\documentclass[journal=jctcce,manuscript=article]{achemso}
\usepackage[T1]{fontenc} % Use modern font encodings
\usepackage{times}
\usepackage{amsmath,amssymb}
\usepackage{amsthm,verbatim, bbm, color, graphicx, geometry, enumerate}
\newcommand{\ua}{\underline{a}}
\newcommand{\ub}{\underline{b}}
\newcommand{\uc}{\underline{\xi}}

\newcommand{\uw}{\underline{w}}

\newcommand{\ue}{\underline{e}}

\newcommand{\uz}{\underline{z}}

\newcommand{\uzz}{\underline{0}}
\newcommand{\uk}{\underline{k}}

\newcommand{\ii}{{\rm i}}
\newcommand{\dd}{\mathsf{diag}(a^2,b^2,c^2)}
\newcommand{\OO}{\mathbf{\Omega}}
\newcommand{\oo}{\omega}

\newcommand{\PP}{\mathcal{P}}
\newcommand{\A}{\mathcal{A}}
\newcommand{\U}{\mathcal{U}}
\newcommand{\Z}{\mathbb{Z}}

\newcommand{\R}{\mathbb{R}}
\newcommand{\C}{\mathbb{C}}
\newcommand{\ie}{\textit{i.e. }}

\newcommand*{\ratthe}[2]{\vartheta \displaystyle \left[ {#1} \atop {#2}\right]}

\author{Emanuele Maggio}
\email{emanuele.maggio@gmail.com}
\affiliation[Scuola Superiore Meridionale]
{Mathematical and Physical Sciences for Advanced Materials and Technologies (MPHS) Cluster, Scuola Superiore Meridionale, largo San Marcellino, 10, 80127 Naples, Italy}

\title{Berry Phase of Bloch States through Modular Symmetries}

\begin{document}

\begin{abstract}
	The theoretical identification of crystalline topological materials has enjoyed sustained success in simplified materials models, often by singling out discrete symmetry operations protecting the topological phase.
	When band structure calculations of realistic materials are considered, complications often arise owing to the requirement of a consistent gauge in the Brillouin zone, or down to the fineness of its sampling.
	Yet, the Berry phase, acting as topological label, encodes geometrical properties of the system, and it should be easily accessible.
	Here, an expression for the Berry phase is obtained, thanks to analytical Bloch states constructed from an infinite series of $s$-type Gaussian orbitals.
	Two contributions in the Berry phase are identified, with one having an immediate geometric interpretation, being equal to the Zak phase.
	Eigenvalues of a modular symmetry, considered here for the first time in the context of crystalline solid state systems, are put in correspondence with the Zak phase: modular symmetries allow to define a non-trivial action for the spatial inversion also when the system does not have an inversion centre, as for the considered case of space group no. 22 ($F222$), which is known to host symmetry equivalent Bloch states distinguishable by their Berry phase.
\end{abstract}

%%%%%%%%%%%%%%%%%%%%%%%%%%%%%%%%%%%%%%%%%%%%%%%%
%% Introduction
%%%%%%%%%%%%%%%%%%%%%%%%%%%%%%%%%%%%%%%%%%%%%%%%
\section{Introduction}
%*) $\Z_2$ invariant has a simple expression for systems with inversion symmetry $\PP$
	The facile identification of crystalline topological systems has taken advantage of symmetry considerations\cite{Slager2013,Po2017,Chang2018,Song2018b,Yoshida2019a,Zhao2020,Kim2019b,Kim2020a,Wang2023NatPhys}, 
	with topological invariants being defined on the score of symmetry eigenvalue conservation.
	The prime example is the $\Z_2$ invariant, that for centrosymmetric systems takes a particularly straightforward formulation\cite{Fu2007a,Fu2007} and which is related to other integral topological invariants\cite{Fu2006a,Wang2010b}.
	The characterisation of crystalline systems without inversion symmetry in the presence of time-reversal symmetry is more convoluted, as there is no easy formula that allows to distinguish topologically non-trivial band structures \cite{Fu2011}.
	 This has prompted the proliferation of computational schemes  either relying on other symmetries protecting the topological phase \cite{Po2017,Kruthoff2017,Song2018b,Zhang2018,Tang2018a,Xiao2021}, or based on a particular representation of the materials' electronic structure.
	In the latter category one finds methods explicitly based on a localised Wannier representation \cite{Soluyanov2011,Yu2011,WannierTools,Gresch2017} or elementary band decompositions \cite{Zak1981,Michel1999,Bradlyn2017}.
	
	The implementation of either method is not without challenges: to track the evolution of the Wannier charge centre, for instance, it is necessary to finely sample the Brillouin zone as the dependence on the wavevector can only be computed numerically, whereas the framework of topological quantum chemistry \cite{Bradlyn2017,Cano2020} requires a smooth gauge along the band dispersion in order to consistently evaluate the transformation character of Bloch states (BSs) under symmetry operations of the local symmetry group for different wavevectors.
	Furthermore, the occurrence of fragile topological bands  \cite{Po2018}, for which a Wannier representation, originally obstructed, can be realised by supplying additional degrees of freedom, leads to a determination of the topological nature of the band structure that  is not  stable against the addition of a topologically trivial set of bands.
	This phenomenon is at odds with the mathematical properties of the (first and second) Stiefel-Whitney classes which indicate an obstruction to orientability in the case of spinless, time-reversal invariant fermions or an obstruction to defining a spin structure as a function of the wavevector.
	The first and second Stiefel-Whitney numbers have been related to the Berry phase and the Wilson loop, respectively \cite{Ahn2018,Bouhon2019}.

	In this article I am considering time-reversal invariant spinless BSs for which the calculation of the Berry phase is sufficient to characterise topologically inequivalent electronic bands.
	The choice of neglecting the spin-orbit coupling allows to sidestep the construction of the double cover of the local symmetry group and to endow the action of symmetry operations with a direct physical meaning that is likely to carry on to the case of non-spin polarised solutions, in the presence of spin-orbit coupling.
	In particular, I am going to derive an analytical expression for the Berry phase 
	and apply it to the case of space group (SG) no. 22, which is a classic example \cite{Zak1991,Michel1992} where symmetry-equivalent BSs are not physically equivalent, \ie they can only be distinguished by their Berry phase.
	The analytical calculation of the Berry phase is made possible by a recent development \cite{Maggio2025} where the BS is expressed as an analytic (in fact entire) complex function and retains the local information of the Gaussian type orbital (GTO) from which it is formed.
	In this work only $s$-type GTOs are considered, with the resulting Berry phase yielding a transparent geometric interpretation, whereas for of $p$- or $d$-GTOs a more detailed analysis is required owing to the Berry phase dependence on the  Gaussian's broadening. 
	The dependence of topological invariants on a toggle that intuitively describes  electron localisation is best interpreted within the theory of the insulating state in crystalline media \cite{Kohn1964,Resta1999,Resta2011} and it is the topic of forthcoming research.
	
	Alternative approaches to the classification of topological materials include their characterisation in terms of symmetry indicators \cite{Po2020,Song2018b}, which, unsurprisingly, turn out to be related to Berry phases (or differences thereof) along specific directions in the unit cell \cite{Song2018a}.
	Unfortunately, the identification of topological crystalline materials can not rely exclusively on symmetry indicators \cite{Bouhon2020}, since the mapping from symmetry indicators to topological invariants in not bijective.
	In this article, a connection is established  between the group of modular transformations of the BSs, whose action has never been considered before, with the expression for the Berry phase.
	The geometrical interpretation that can thus be attributed to the Berry phase is subject to a symmetry enforced cancellation of one of its two components identified in  Eq. \ref{eq:Berryphase} in the next section.
	Whenever such cancellation occurs, the evaluation of the Berry phase reduces, in fact, to that of a symmetry eigenvalue, overcoming the limitations mentioned above for other established BS representations, where the direct evaluation of the corresponding integral is required.
	On the other hand, the Riemann BS could be envisaged as a basis set for realistic materials band structure calculations, where the (mean-field) solution of the crystalline Hamiltonian would be given by a superposition of symmetry-adapted BS, as suggested in Ref. \cite{Strinati1978}.
	%interpolate as a function of the wavevector between the 
	%By following the prescriptions in the seminal work of Strinati \cite{Strinati1978}, the present BS representation could be envisaged as a basis set for mean-field level calculations of realistic materials; 
	In such a context, the results presented here can be interpreted as the limiting case of an isolated band with no admixture of BSs.

%%%%%%%%%%%%%%%%%%%%%%%%%%%%%%%%%%%%%%%%%%%%%%%%
\section{Results}
%%%%%%%%%%%%%%%%%%%%%%%%%%%%%%%%%%%%%%%%%%%%%%%%

%===== Bloch states =====
\subsection{Analytical evaluation of the Berry phase for Riemann Bloch states}
% *) Bloch states generated by resummation of GTOs, factorisation based on the form of $\Omega$
	Riemann $\vartheta$ functions have recently been  proposed \cite{Maggio2025} to express BSs arising from GTOs centred at a given Wyckoff position (WP), indicated by the symbol $\uw$; the resulting BSs are immediately associated with a label conveying local information concerning the  orbital's angular momentum ($s, p, d, \dots$) and its location, by construction.
	The analytical expression for BSs depends on the complex variable $\uz = \uk - \OO \uc$ in which the wavevector $\uk$ and the spatial coordinate $\uc$ are both referred to the conventional unit cell, $\U$. % in reciprocal and direct space respectively. 
	For an $s$-GTO the resulting Bloch state is:
 \begin{align}
 \label{eq:BS}
 \phi_{\uk}(\uc|\uw,\OO) = N_{\uk} \, e^{\ii \pi \uc \cdot \OO \uc} \, e^{-2 \ii \pi \uw \cdot \uk} \, \ratthe \uw \uzz (\uk - \OO \uc | \OO),
 \end{align}
 where the Riemann $\vartheta$ function with characteristics is defined in Sec. \ref{ssec:overlap} and $N_{\uk}$ is a (wavevector dependent) normalisation constant.
 For notational convenience, the dependence on the Gaussian broadening $\beta$ is implicit in the period matrix $\OO$. 
	To derive an analytical expression for the Berry phase and connection, the latter is assumed diagonal, $\ie$ $\OO = \tau \mathsf{diag}(a^2, b^2, c^2)$ with $a, b, c$ lattice constants and $\tau=\tfrac{\ii \beta}{\pi}$, in which case the Riemann $\vartheta$ function reduces to the product of one-dimensional Jacobi functions (defined in Sec. \ref{ssec:Jac}) in each coordinate.
	%{\bf[say what happens when the period matrix does not factorise]}
	This assumption conspicously leaves out SGs with triclinic, monoclinic and hexagonal Bravais lattices for which, to the best of a my knowledge, only a numerical approach remains viable within the present framework.
	In the following, Riemann and Jacobi functions will be distinguished by their vector or scalar input variables, respectively, although different notations for the two functions are common in the literature \cite{DHoker2025,Diamond2005,NISThandbook}.

	Factorisation of the BS and orthogonality of the conventional crystal coordinates imply that for the normalisation constant $N_{\uk}$ one has:
 $ N^2_{\uk} = |T| \frac{1}{S_{\uk}} = |T| \frac{1}{S_{k^1}S_{k^2}S_{k^3}}$, %where $T$ is the matrix containing the Bravais lattice basis vectors.  %%define T and N_k...
 which derives from the normalisation condition $N^2_{\uk} \langle \tilde{\phi}_{\uk} | \tilde{\phi}_{\uk}\rangle =1$, with $\tilde{\phi}$ being the unnormalised BS and the bracket $\langle \bullet|\bullet \rangle$ indicating spatial integration over $\U$. 
	In the following the determinant of the matrix $T$ containing the Bravais lattice basis vectors is omitted in the expression for the one-dimensional component $N_{k^\mu}=(S_{k^\mu})^{-\tfrac{1}{2}}$.
	The directional superscript $\mu$ will be dropped in the following, if this is causes no misunderstanding.
 
% *) overlap as function of the wavevector (implication for reality of Berry connection)
	The expression for the overlap $S_k$ of a $\vartheta$ function with respect to one coordinate (with lattice constant $a$ along that direction), derived in Sec. \ref{ssec:overlap}, is given by another $\vartheta$ function with scaled period $\oo = \tfrac{1}{2} \tau a^2$:
 \begin{align}
 \label{eq:ovrl}
 S_k=\sqrt{\frac{\pi}{2\beta}} \, \vartheta(k|\tfrac{1}{2} \tau a^2).
 \end{align}
	Two considerations emerge from Eq. \ref{eq:ovrl} above: first, the explicit dependence on the wavevector implies that $\nabla_{\uk} \langle \phi_{\uk} | \phi_{\uk}\rangle \neq 0$ and consequently the Berry connection $\A_{\uk} \equiv \ii \left\langle u_{\uk}| \nabla_{\uk} \, u_{\uk}\right\rangle$ is not purely real on general grounds (\textit{vide infra} for the definition of $u_{\uk}$, the BS's periodic component). 
	In Sec. \ref{ssec:Jac} it is shown that for $s$-type BSs the  imaginary contribution vanishes, however I can anticipate that careful analysis is necessary for BSs of higher orbital momentum, which will be reported elsewhere.
	Second, the BS is well defined as the normalisation constant does not vanish anywhere in the Brillouin zone: this descends from the zeros of a $\vartheta$ function being located at $\tfrac{1}{2} + \tfrac{1}{2}\oo + \Lambda$, where $\Lambda$ is the complex lattice generated by $(1,\oo)$ \cite{Mumford1983,Diamond2005,DHoker2025}.

%===== Berry connection and Berry phase =====
	To evaluate the Berry connection, the periodic component of the BS $u_{\uk}= e^{-2\ii \pi \uc \cdot \uk} \, \phi_{\uk}$ is introduced and the evaluation of its derivative $\tfrac{\partial}{\partial k^\mu} u_{\uk}$ is reported in the Supporting Information.
	A crucial ingredient is the derivative of the normalisation constant:
	\begin{align}
	\label{eq:dNdk}
	\frac{\partial}{\partial k} N_k = -\frac{1}{2} N_k \, \frac{\partial}{\partial k} \left( \ln S_k \right) = -\frac{1}{2} N_k \frac{\vartheta '(k|\tfrac{1}{2}\tau a^2)}{\vartheta(k|\tfrac{1}{2}\tau a^2)},
	\end{align}
	which leads to the presence of two components in the Berry connection, namely the "dispersive" and "geometrical" component:
 \begin{align}
 \label{eq:Connct}
 \left\langle u_{\uk} \bigg| \frac{\partial}{\partial k^\mu} u_{\uk}\right\rangle = -\frac{1}{2} \left[ \frac{\vartheta '(k^\mu|\tfrac{1}{2}\tau a^2)}{\vartheta(k^\mu|\tfrac{1}{2}\tau a^2)} + 4\pi \ii (\ue^\mu , \uw) \right]. 
 \end{align}
	The geometrical component consists of the inner product $(\ue^\mu,\uw)$ between the direction with respect to which the derivative is taken and the Wyckoff position, whereas the dispersive component retains a dependence on the wavevector (hence the name) and it is given by the logarithmic derivative of the overlap function.
	For $s$-GTOs the latter returns the ratio on the right hand side in Eq. \ref{eq:dNdk}.
 
	From Eq. \ref{eq:Connct} it is immediately apparent that the Berry connection is a closed differential form (its curl evaluates to zero), and, with its integration in the Brillouin zone being path independent, it follows that the integral over a homotopically trivial loop is identically zero.
	Therefore, the Berry phase can be  meaningfully computed by considering as endpoints of the integration path (high symmetry) wavevectors on opposite faces of the Brillouin zone (that are identified under periodic boundary conditions), or an open path in the Brillouin zone, which thanks to the gauge enforced by the BSs will unambiguously specify the Berry phase, but will  
	not be invariant under SG symmetries in general.
	Also, with the Berry curvature $\mathcal{F} = \nabla_{\uk} \times \mathcal{A}_{\uk}$ being equal to zero, it follows that the invariant introduced in Ref. \cite{Cano2022} reduces in this gauge to the exponential of the Berry phase.
 
%===== expression for the Berry phase =====
	The translational invariance of the $\vartheta$ functions with respect to the wavevector allows one to fix the integration interval as $[0,m]$.
	If the integral is evaluated along the direction $\ue^\mu$ the Berry phase is given by: 
 \begin{align}
 \label{eq:Berryphase}
 \ii \gamma(m) = \int_0^m dk^\mu \; \frac{1}{2} \left[\frac{\vartheta '(k^\mu|\tfrac{1}{2}\tau a^2)}{\vartheta(k^\mu|\tfrac{1}{2}\tau a^2)} + 4\pi \ii (\ue^\mu \cdot \uw)  \right],
 \end{align}
where the geometric component evaluates straightforwardly to $2\pi \ii m (\ue^\mu \cdot \uw)$ and the dispersive component integrates to zero for the example considered in the next section.
	Note that if the upper integration limit is not an integer, as it might happen if an open path in reciprocal space is considered for primitive Bravais lattices, the dispersive component of the Berry connection does not vanish in general; it is this component that introduces a dependence on the GTO broadening when $p$- and $d$- orbitals partake in the formation of the BS and which can help discriminate between topologically trivial and non-trivial electronic bands for a given crystal structure.

\subsection{Application to symmetry-indistinguishable bands in SG no. 22 (F222)}
%% application to the case of bands labelled by different Wyckoff positions that can not be distingushed by symmetry
	To showcase that the action of modular symmetries can be successfully exploited to connote the distinct topological nature of electronic bands, the case of SG no. 22 ($F222$) is now considered.
	This space group has emerged as an early exception to the paradigm of a symmetry based classification of band representations, since the symmetry equivalent WPs give out physically non-equivalent band representations \cite{Bacry1988}, as they are associated to different values of the Berry phase \cite{Michel1992}.

\begin{figure}
\includegraphics[width=180mm]{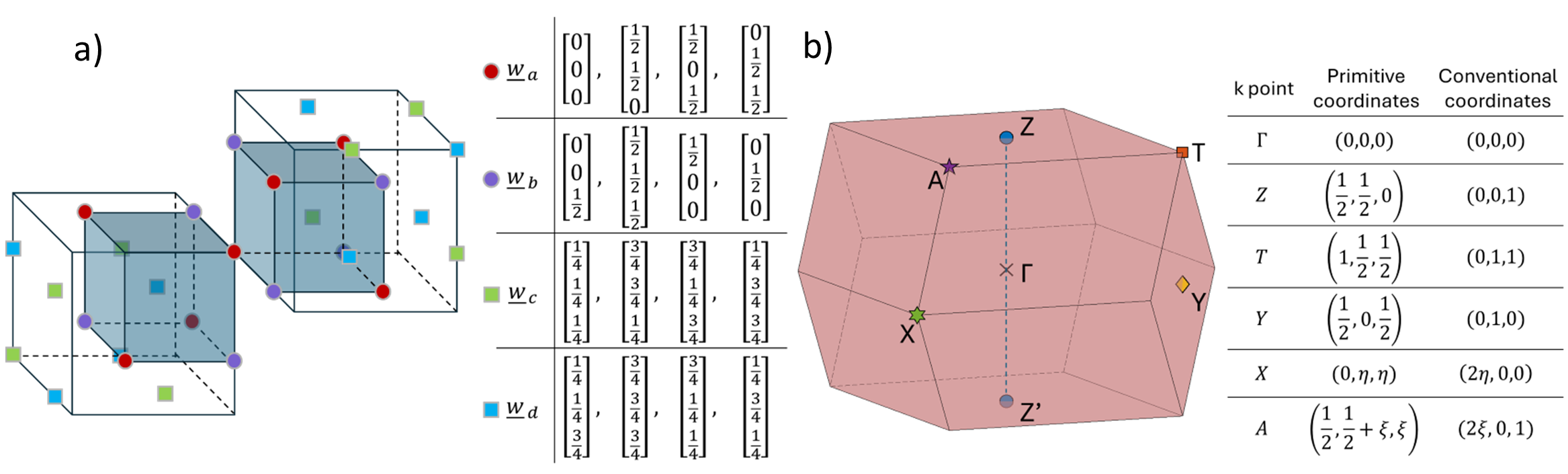}
\caption{High symmetry points in direct (panel a) and reciprocal (panel b) space for space group $F222$ no. 22. In panel a) the Wyckoff positions $\uw_a$ to $\uw_d$ are shown in the unit cell and in its image under inversion symmetry.  The cubes with edge length $\tfrac{1}{2}$ (shaded faces) and $\tfrac{3}{4}$ (transparent faces) have been included as a guide to the eye. Wyckoff coordinates in the unit cell for each position are indicated. In panel b) the Brillouin zone corresponding to the choice of lattice parameters $\tfrac{1}{a^2} = \tfrac{1}{b^2} + \tfrac{1}{c^2}$ is shown with lattice-dependent high-symmetry wavevectors coordinates $\eta = \tfrac{1}{4}(1+(\tfrac{a}{b})^2) + (\tfrac{a}{c})^2$ and $\xi = \tfrac{1}{4}(1+(\tfrac{a}{b})^2) - (\tfrac{a}{c})^2$ \cite{Maggio2023}. The integration path for the calculation of the Berry phase carried out in the text is indicated by a dashed line.}
\label{fig:SG22}
\end{figure}

	The WPs considered are indicated by the symbol $\uw_\alpha$ with $\alpha \in \{a, b, c, d \}$; their multiplicity is four, \ie under the action of SG operations there are four equivalent coordinates for each WP, belonging to the same orbit, indicated by $\{\uw_\alpha\}$.
	The equivalent coordinates in each WP are reported in Fig. \ref{fig:SG22} panel (a).
	SG elements that keep each coordinate fixed form the stabiliser group, $\mathsf{Stab}(\uw_\alpha)$, of the WP. 
	 For the four positions $a, b, c, d$ we have that the stabilisers are all isomorphic and pairwise equal, \ie $\mathsf{Stab}(\uw_\alpha) \cong (\Z_2 \times \Z_2)$ and $\mathsf{Stab}(\uw_a) = \mathsf{Stab}(\uw_b)$, $\mathsf{Stab}(\uw_c)=\mathsf{Stab}(\uw_d)$.
	Since the intersection of WPs' orbits is empty, it follows that the union of their stabilisers can not be a finite subgroup of the space group, so it must contain a lattice translation \cite{Fuksa1994}. 
	This property characterises maximal WPs, and the corresponding band representations are elementary \cite{Michel2001}, owing to the stabilisers being Abelian. 
	In such a case, the global band structure connectivity is determined by group-theoretical arguments (the so-called compatibility relations) if one knows which WPs partake in the material's electronic structure and the BSs' transformation properties are thus fixed \cite{Zeiner2000}. 
	For the SG considered here, symmetry operations can not distinguish BSs pertaining to WPs in either pair $(a,b)$ or $(c,d)$, then the determination of the band structure connectivity requires additional, topological, labels, provided by Berry phases.
	
	%say how you compute the Berry phase
	The Berry phase is computed according to Eq. \ref{eq:Berryphase}, with the choice of $\ue^\mu=\ue^3$ along the $\uk_Z$ direction in  reciprocal space, shown by a dashed line in Fig. \ref{fig:SG22} panel (b); the translational invariance of the $\vartheta$ function with respect to the wavevector allows to fix the lower integration limit at the $\Gamma$ point and by setting $m=2,1$, respectively, one evaluates the integral in Eq. \ref{eq:Berryphase} along a closed loop or along the open path $\Gamma-Z$, which reproduces previous results \cite{Michel1992} reported in Tab. \ref{tbl:results}.

% introduction of modular symmetry (modular parity?)
	Next, I am considering the action of the inversion symmetry $\PP$.
	Even though the space group is not centrosymmetric, the inversion is an element of the modular group (defined in the Supporting Information) acting on $\vartheta$ functions and modular forms. 
	The symmetry eigenvalue for the modular component of the BS can, therefore, be estimated and it is found to coincide with  $e^{\ii \gamma}$,  thus providing an unexpected link between a topological invariant and a class of modular symmetries.
	
% action of the inversion symmetry on the different sublattices generated by the WPs 
	First, let me examine the Wyckoff positions $\uw_\alpha$ that are shown in Fig. \ref{fig:SG22} panel (a) within the unit cell and in its image generated by the inversion symmetry $\PP$.
	Since $\PP$ is not a space group operation, the corresponding orbit of $\uw_\alpha$ (\ie $\PP \uw_\alpha$)  will not belong, in general, to the sublattice generated by $\{ \uw_\alpha \}$ under lattice translations.
	However, coordinates in the pair $(a,b)$ are fixed under $\PP$, whose  action, instead, swaps coordinates belonging to the pair $(c,d)$; hence the sublattices generated by $\{\uw_c \}$ and $\{\uw_d \}$ are identified under inversion.
	This behaviour is shared by BSs generated by GTOs located at WP coordinates $\{ \uw_\alpha \}$ (indicated in the following with the notation BSs @ $\{\uw_\alpha\}$) and it is analysed in Sec. SI.1 of the Supporting Information, where it is shown that the inversion symmetry applied to the BSs returns a trivial eigenvalue, which therefore cannot serve as a topological label for the dispersion involved; the opposite is true when $\PP$ is applied to the \textit{modular} component of the BSs.
	
%%%%%%%%%%%%%%%%%%%%%%%%%%%%%%%%%%%%
%% Figure 1
%%%%%%%%%%%%%%%%%%%%%%%%%%%%%%%%%%%%

	To prove the claim above, recall that the  action on the $\vartheta$ functions evaluates as $\PP \left[ \ratthe \uw \uk (\uz|\OO) \right] = \ratthe \uw \uk (-\uz|\OO) = \ratthe {-\uw} {-\uk} (\uz|\OO)$ \cite{Mumford1983,Maggio2025}.
	For the pair $(a,b)$ the coordinates $\uw_\alpha$ have integer or half-integer entries and the corresponding BSs are eigenfunctions of the inversion centre, with eigenvalue $e_*(\uw_\alpha,\uk)=(-1)^{4 \, \uw_\alpha \cdot \uk}$ for any $\uk \in \tfrac{1}{2}\Z^3$; the parity eigenvalue evaluates to $+1$ at the $Z-$point, as reported in Tbl. \ref{tbl:results}.
	For the $\vartheta$ functions with characteristics $\uw_c$ or $\uw_d$, on the other hand, one has that the action of the inversion symmetry matches that on the corresponding sublattices generated by the WPs: a direct calculation shows that $\ratthe {-\uw_c} {-\uk} (\uz|\OO) = \ratthe {\uw_d - \underline{1}} {\uk -2\uk} (\uz|\OO) = (-1)^{4 \, \uw_d\cdot\uk}\ratthe {\uw_d} {\uk}(\uz|\OO) = e_*(\uw_d,\uk) \, \ratthe {\uw_d} {\uk}(\uz|\OO)$, for any $\uk-$point with integer or half-integer coordinates.
	 Hence, the $\vartheta$ functions retain the action of the inversion symmetry on the two sublattices (with WPs $\uw_c$ and $\uw_d$ being swapped), and their symmetry eigenvalues are not affected, taking value $-1$ in either case.
	
	The Berry phase evaluated along the closed loop $Z'-\Gamma-Z$ can be obtained by setting $m =2$ in Eq. \ref{eq:Berryphase} and the corresponding numerical values are reported in Tab. \ref{tbl:results}, with $e^{\ii \gamma}$ found to coincide with the parity eigenvalues just computed.
	In order to discriminate between the topologically distinct BSs, the Berry phase has to be evaluated on the open path $\Gamma-Z$ (corresponding to $m=1$ in Eq. \ref{eq:Berryphase}), or along any reciprocal lattice vector\cite{Michel1992}, the parity of the representative $\vartheta$ function evaluated at the "barycenter" of the dispersion (\ie at the midpoint of the integration interval) again matches the computed values for the Berry phase.
	Next I am going to contrast the derivation above with a 
	previously reported  expression\cite{Zak1989} obtained in a less general setting and compare the BSs features that are encoded in the Berry phase.

%%%%%%%%%%%%%%%%%%%%%%%%%%%%%%%%%%%%
%% Table results
%%%%%%%%%%%%%%%%%%%%%%%%%%%%%%%%%%%%
\begin{table}
\caption{Berry phase $\gamma$ and parity eigenvalue computed for the constituent $\vartheta$ functions of $s$-BSs located at the representative Wyckoff position coordinate $\uw_\alpha$ (reported in the first column of the list of Fig. \ref{fig:SG22}). The parity eigenvalue is evaluated with $\beta = \alpha, \, j = 1$ for $\alpha=a,b$ and $\beta = d,c \, j=2$ for $\alpha=c,d$ respectively.}
\label{tbl:results}
\begin{tabular}{c|ccc|ccc}
%\hline
$\alpha$	& $\gamma(2)$ & $e^{\ii \gamma}$ & $e_*(\uw_\beta^{(j)},\uk_Z)$ & $\gamma(1)$ & $e^{\ii \gamma(1)}$ & $e_*(\uw_\beta^{(j)},\tfrac{1}{2}\uk_Z)$\\
\hline
$a$ & $0$ & $\phantom{-}1$ & $\phantom{-}1$ & $\phantom{-}0$ & $\phantom{-}1$& $\phantom{-}1$\\	
$b$ & $2\pi$ & $\phantom{-}1$ & $\phantom{-}1$ & $\phantom{-}\pi$ & $-1$& $-1$\\	
$c$ & $\pi$ & $-1$ & $-1$ & $\phantom{-}\frac{\pi}{2}$ &$\phantom{-}\ii$&$\phantom{-}\ii$\\	
$d$ & $3\pi$ & $-1$ & $-1$ & $-\frac{\pi}{2}$ & $-\ii$ & $-\ii$\\	
\hline
\end{tabular}
\end{table}

%%%%%%%%%%%%%%%%%%%%%%%%%%%%%%%%%%%%%%%%%%%%%%%%
\section{Discussion}
%%%%%%%%%%%%%%%%%%%%%%%%%%%%%%%%%%%%%%%%%%%%%%%%

\label{sec:Dsc}
	
\begin{figure}[h]
\includegraphics[width=180mm]{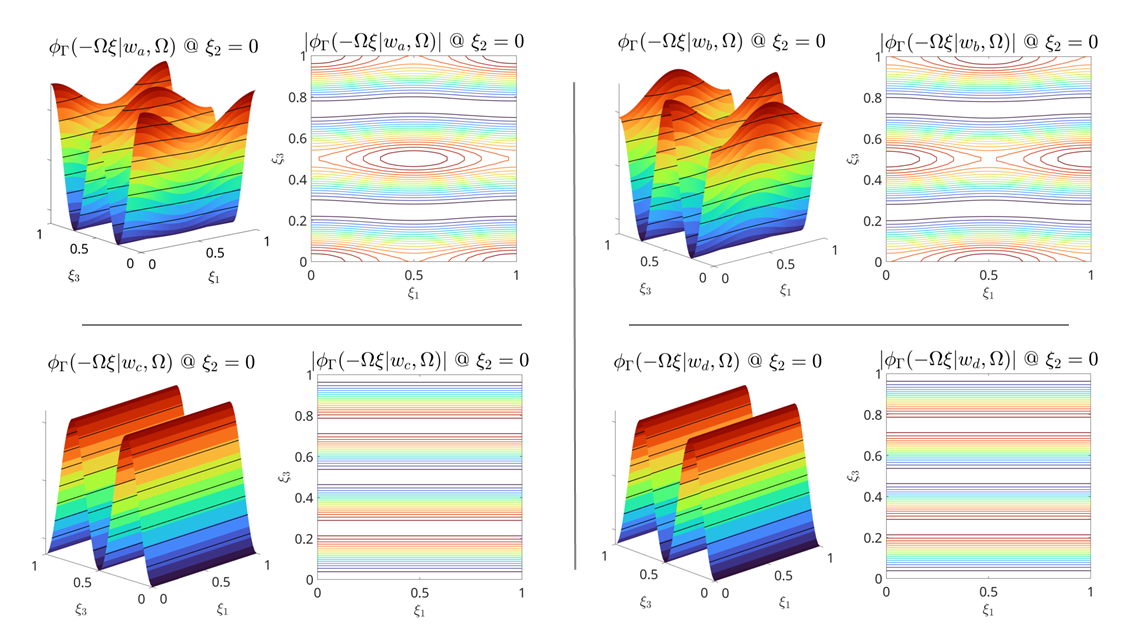}
\caption{Bloch states @ $\{ \uw_\alpha\}, \, \alpha \in \{a,b,c,d\}$. Each panel shows the real part (left) and the absolute value of the BS in the plane $\xi_2 = 0$, the imaginary part of the Bloch states is zero at the $\Gamma-$point.}
\label{fig:BSs}
\end{figure}

%Interpretation of results in terms of symmetries of the Bloch states
	The BSs considered are the result of a Bloch-Floquet transform \cite{Kuchment1982,Dirl1996a} of a symmetric superposition of $s$-GTOs centred at the coordinates in $\{\uw_\alpha\}, \, \alpha \in \{a,b,c,d\}$.
	At the $\Gamma-$point all BSs transform like the trivial representation (see Tbl. S1 in the Supporting Information), hence they are all invariant under SG symmetries.
	In Fig. \ref{fig:BSs} the section of the BSs along the plane $\xi_2=0$ are reproduced, with BS@$\{\uw_a\}$ and BS@$\{\uw_b\}$ at the top and BS@$\{\uw_c \}$, BS@$\{\uw_d\}$ in the bottom row.
	The relative shift of the BSs associated to the two WPs pairs is shown in the pictures and stems from the same property of the constituent $\vartheta$ functions,  
	reflected by their parity eigenvalue evaluated at the $Z-$point (note that the BSs at $\Gamma$ and $Z$ are identical) %, owing to the invariance of the BS with respect to integer translations of the wavevector),
	 and consistently, by the $\gamma(2)$ values.
	
	To get a more granular distinction between the states one has to evaluate $\gamma(1)$ (in agreement with previous studies \cite{Michel1992,Cano2022}) whose value uniquely identifies each of the BSs considered, starting with the pair $(a,b)$ in Fig.\ref{fig:BSs}, for which the Berry phase $\gamma(1)$ is a real number.
	One readily identifies the translation by $\tfrac{1}{2} \ue^3$ as the symmetry breaking transformation (it is not an element of the SG) able to map BS@$\{\uw_a\}$ to BS@$\{\uw_b\}$.
	The expression for the BSs is invariant by shifting the unit cell origin (as reported in Sec. \ref{ssec:overlap} and previously shown\cite{Maggio2025}) hence the relative phase difference along the $\xi_1$ and $\xi_2$ directions (the latter shown in Fig. S1 in the Supporting Information) connotates two physically inequivalent BSs.
	In contrast, for the case of BS@$\{\uw_c\}$ and BS@$\{\uw_d\}$ the exponential of the Berry phase takes purely imaginary values and the two states, in fact, coincide.	

%Conjecture
	A plausible explanation for the observed behaviour involves the relationship that has been established in this article between the Berry phase and system's symmetries.
	In general, the BSs form a basis for the symmetry invariant manifolds associated to irreducible representations of the local symmetry group of the wavevector. 
	For the case at hand, all irreducible representations have real characters, and one is led to conjecture that the reality of the manifolds does not allow to tell apart (by the action of symmetry transformations) states that are associated with imaginary values taken by the Berry phase.
	It would be interesting to test this conjecture on the many pairs of (physically) inequivalent WPs identified in Ref. \cite{Cano2022} and verify if for the case of complex characters, there is a symmetry operation able to distinguish BSs associated to imaginary values of the Berry phase.

%Concluding remarks
	As a final remark, note that the integrated expression for the geometric component in Eq. \ref{eq:Berryphase} is identical to the one derived by Zak more than a quater of a century ago \cite{Zak1989}; in order to make a connection to the WP, in that work, the presence of the inversion centre in a one-dimensional material model was assumed.
	Here, the derivation of the expression for the Berry phase is more general and rests only on the factorisation of the BS in terms of Jacobi $\vartheta$ functions and on their peculiar property for the overlap integral, which is just another $\vartheta$ function with scaled period. 
	The resulting expression for the Berry phase naturally differentiates the contribution of the dispersive component and the one stemming from the geometry of the crystalline system, thus reconciling the interpretation of Zak with a more fundamental electronic structure representation. 
	For BSs constructed with $s$-GTOs considered in this work, this geometric contribution completely characterises the Berry phase whenever the dispersive component vanishes by symmetry, \ie when the Berry phase is evaluated with respect to integer wavevector coordinates occurring in non-primitive Bravais lattices.
	In this case the Berry phase is protected by a symmetry of the electronic state, which acts on the modular component of the BS.

%%%%%%%%%%%%%%%%%%%%%%%%%%%%%%%%%%%%
%% Figure 2
%%%%%%%%%%%%%%%%%%%%%%%%%%%%%%%%%%%%

\section{Methods}
\label{sec:Mth}
	The numerical evaluation of $\vartheta$ functions has been carried out as in Ref.\cite{Maggio2025} ,  omitting the normalisation constant for the GTOs.
	The approximation method banks on the pointwise approximation introduced in Ref.\cite{Deconinck2003} , which allows to set the numerical error in the evaluation of the $\vartheta$ function as an input parameter, here set to $10^{-8}$.
	The analytical expression for the overlap integral has been compared to the result of numerical integration using the built in trapezoid method implemented in \textsc{Matlab} \cite{Matlab2024}: setting the lattice constants to the numerical values $a=1, \, b= \tfrac{5}{4}, \, c= \tfrac{5}{3}$, the numerical approximation $S^{(N)}$ to $S$ already converges to the analytical value with an absolute error $\left| S^{(N)} - S \right|\approx 10^{-11}$ for $N \geq 5$, for the BS $\tilde{\phi}_\Gamma(\uc|\uw_a, \tau \dd)$, with $\tau=\tfrac{2\ii}{\pi}$.
	
%	- transformation properties of the Bloch states (link to SI for the character tables)
	The assignment of a BS to an irreducible representation of the local symmetry group of the wavevector is carried out by evaluating the ratio $\tfrac{F[\phi]}{\phi}$, where $F$ is the SG operation \cite{Maggio2025}, the result is compared with the character table reported in the Supporting Information.

%% statement of how this section is organised
	In the rest of this section I am reporting the derivation of the salient results presented in the main text, together with the necessary definitions, starting with the integral of the Berry connection that is somewhat easier to derive.
\subsection{Integration of the dispersive component of the Berry connection}
\label{ssec:Jac}
Define the Jacobi $\vartheta$ function as follows:
\begin{align}
\label{eq:theta}
\vartheta(z|\oo) = \sum_{n=-\infty}^{+\infty} e^{\ii \pi \oo n^2} \, e^{2 \pi \ii n z} = \sum_{n=-\infty}^{+\infty} v^{2n} q^{n^2},
\end{align}
with $z \in \C$ the argument, the parameter $\oo \in \C, \, \Im \{\oo\} >0$ being the period and $q=e^{\ii \pi \oo}$ typically referred to as the nome \cite{NISThandbook}. The Jacobi triple product states (see Ref. \cite{DHoker2025} Eq. 2.6.24 and Ref.\cite{NISThandbook} Eq. 20.5.9): $ \sum_{n=-\infty}^{+\infty} v^{2n} q^{n^2} = \prod_{n=1}^{+\infty} (1-q^{2n})(1+(v^2 + v^{-2})q^{2n-1} +q^{4n-2})$ which can be rewritten as $q_0 \prod_{n=1}^{+\infty} c_n$, with $q_0=\prod_{n=1}^{+\infty} 1-q^{2n}$ and $c_n(z)=1+2\cos(2\pi z)q^{2n-1} +q^{4n-2}$.
	Then, for the ratio of the function and its derivative we have:
	\begin{align}
	\frac{\vartheta'(z|\oo)}{\vartheta(z|\oo)}=\frac{\tfrac{d}{dz}\prod c_n(z)}{\prod c_n(z)}=\sum_{n=1}^{+\infty}\frac{c_n'}{c_n}=-4\pi \sin(2\pi z) \sum_n \frac{q^{2n-1}}{1+2\cos(2\pi z)q^{2n-1}+q^{4n-2}}
	\end{align}
	and the integral over the real part of the complex variable $\Re\{z\}=k$
	\begin{align*}
	\int_0^m dk \frac{\vartheta'(k|\oo)}{\vartheta(k|\oo)}= 2 \sum_n (-q)^{2n-1} \, 2\pi \int_0^m dk \frac{\sin(2\pi k)}{1-2\cos(2\pi k) (-q)^{2n-1} + \left( (-q)^{2n-1} \right)^2}
	\end{align*}
	vanishes identically, since for a given value of $n$ one has:
	\begin{align}
	2\pi \int_0^m dk \frac{\sin(2\pi k)}{1 -2a\cos(2\pi k) +a^2} = \int_0^{2\pi m} dx \frac{\sin(x)}{1-2a\cos(x)+a^2}=\frac{1}{2a}\int_{(1-a)^2}^{(1-a)^2} \frac{du}{u}=0
	\end{align}
	owing to the integration limits of the transformed integral coinciding for $m \in \Z$.

\subsection{Overlap integral for s-GTO BSs}
\label{ssec:overlap}
The Riemann $\vartheta$ function with characteristics $\ua, \ub \in \R^3$ is defined as \cite{Mumford1983,NISThandbook}:
\begin{align}
\label{eq:ratthe}
\ratthe{\ua}{\ub}(\uz|\OO) = e^{\ii \pi \ua \cdot \OO \ua} \, e^{2\pi \ii \ua \cdot (\uz + \ub)} \, \vartheta(\uz + \OO\ua + \ub|\OO),
\end{align}
%closed expression for a 1D overlap integral expression derived on CC-12/CC-14
	with $\uz \in \C^3$ and $\OO$ an element of the Siegel upper-half space, that is $\OO \in \C^{3 \times 3}$ is symmetric and its imaginary part is positive definite.
The exponential terms in Eq. \ref{eq:ratthe} ensure the translational invariance of the BS, which implies that for an $s$-BS one has $\phi_{\uk}(\uc|\uw,\OO) = \phi_{\uk}(\uc-\uw|\uzz,\OO) = \varphi_{\uk}(\uc-\uw|\OO)$.
If the overall BS factorises (as it happens for a diagonal period matrix $\OO$), then it is possible to evaluate the overlap integral analytically along each cartesian direction, with $a$ the corresponding lattice parameter and $W$ the Wyckoff position component in those coordinates.
	The expression for the overlap integral reads:
\begin{align}
S_k(\tau a^2) = \int_0^a dx \; \tilde{\varphi}_{-k} \left(\tfrac{1}{a}(x-W)\big | \tau a^2 \right) \, \tilde{\varphi}_{k} \left(\tfrac{1}{a}(x-W)\big | \tau a^2 \right)
\end{align}
where the reality of the Gaussians allows one to consider the BS at momentum $-k$ in stead of the complex conjugate function \cite{Zeiner1998}.
Inserting the definition in Eq. \ref{eq:theta} one writes the integral as:
\begin{align}
S_k(\tau a^2) & = \sum_{m,n \in \Z} e^{2 \pi \ii k(n-m)} \, e^{-\beta a^2(n^2 + m^2)} \int_0^a dx \; e^{-2\beta (x-W)^2} \, e^{2\beta a (n+m)(x-W)}  \nonumber \\
& = \sum_{m,n \in \Z} e^{2 \pi \ii k(n-m)} \, e^{-\beta a^2(n^2 + m^2)} \, e^{\frac{1}{2} \beta a^2 (m+n)^2} \int_0^a dx \; e^{-2 \beta (x -W - \frac{a}{2}(n+m))^2} \nonumber \\
& = \sum_{m,n \in \Z} e^{2 \pi \ii k(n-m)} \, e^{-\frac{1}{2} \beta a^2 (n-m)^2} \int_0^a dx \; e^{-2 \beta (x -W - \frac{a}{2}(n+m))^2}  \nonumber \\
& = \sum_{s = -\infty}^{+\infty} e^{2 \pi \ii k s} \, e^{-\frac{1}{2} \beta a^2 s^2} \sum_{m =-\infty}^{+\infty} \int_0^a dx \; e^{-2 \beta \left( x - W - am - \frac{1}{2}as \right)^2}
\end{align}
where in going to the last line above we have introduced the summation index $s=n -m$.
Now we continue with a change of variable that shifts the integral: $\tilde{x} = x-am $, thus leading to an integration interval $[-am,a(1-m)]$, which, as the summation index is varied, ends up spanning the whole real axis, thus:
\begin{align}
S_k(\tau a^2) &= \sum_{s = -\infty}^{+\infty} e^{2 \pi \ii k s} \, e^{-\frac{1}{2} \beta a^2 s^2} \; \sum_{m=-\infty}^{+\infty} \int_{-am}^{a(1-m)} d\tilde{x} \, e^{-2\beta\left(\tilde{x} -W -\frac{1}{2}as \right)^2} \nonumber \\
& = \sum_{s = -\infty}^{+\infty} e^{2 \pi \ii k s} \, e^{-\frac{1}{2} \beta a^2 s^2} \; \int_{-\infty}^{+\infty}d\tilde{x} \, e^{-2\beta\left(\tilde{x} -W -\frac{1}{2}as \right)^2} \nonumber \\
& = \sum_{s = -\infty}^{+\infty} e^{2 \pi \ii k s} \, e^{-\frac{1}{2} \beta a^2 s^2} \; \int_{-\infty}^{+\infty} dy \; e^{-2\beta y^2} = \sqrt{\frac{\pi}{2\beta}} \sum_{n =-\infty}^{+\infty} e^{2 \pi \ii k n} \, e^{-\frac{1}{2}\beta a^2 n^2} \nonumber \\
& = \sqrt{\frac{\pi}{2\beta}} \; \vartheta \left( k \bigg | \frac{\ii \beta}{2 \pi} a^2\right) = \sqrt{\frac{\pi}{2\beta}} \; \vartheta \left( k \bigg | \frac{1}{2} \tau a^2\right).
\end{align}

\begin{acknowledgement}
The author sincerely thanks R.-J. Slager for bringing to his attention Refs. \cite{Bouhon2019,Bouhon2020} and G. Calvanese-Strinati for suggesting Ref.\cite{Strinati1978} .
\end{acknowledgement}

\section*{Author contributions}
The article, its data and the theory  have been conceived, elaborated, written and edited by the author.

\begin{suppinfo}

\begin{itemize}
  \item SuppInfo: contains the definition of the Modular group and characterisation of the BS under inversion symmetry, character table and additional derivations.
  \item Competing financial interests: the author declares no competing financial interest.
\end{itemize}

\end{suppinfo}

%\bibliographystyle{achemso}
%\bibliography{Books,Topology,MyArticles,GeneralTheory,GroupTheory,NumericalMethods}
\providecommand{\latin}[1]{#1}
\makeatletter
\providecommand{\doi}
  {\begingroup\let\do\@makeother\dospecials
  \catcode`\{=1 \catcode`\}=2 \doi@aux}
\providecommand{\doi@aux}[1]{\endgroup\texttt{#1}}
\makeatother
\providecommand*\mcitethebibliography{\thebibliography}
\csname @ifundefined\endcsname{endmcitethebibliography}
  {\let\endmcitethebibliography\endthebibliography}{}

\end{document}

\section{Modular group and inversion symmetry characterisation}
The Riemann $\vartheta$ functions in the main text retain a parametric dependence on the period matrix $\OO \in \C^{g \times g}$ which is a purely imaginary matrix with imaginary part being symmetric and positive definite. 
The space of such matrices is denoted by $\mathcal{H}_g$ in the following.
The modular group $Sp(2g,\R)$ is defined as the group of matrices with real entries of the form $\begin{pmatrix}
A & B\\
C & D
\end{pmatrix}$ that preserve the symplectic form $\begin{pmatrix}
0 & \I_g\\
-\I_g & 0
\end{pmatrix} $, that is for which the identity
\begin{align*}
\begin{pmatrix}
A & C\\
B & D
\end{pmatrix}^t \begin{pmatrix}
0 & \I_g\\
-\I_g & 0
\end{pmatrix} \begin{pmatrix}
A & B\\
C & D
\end{pmatrix} = \begin{pmatrix}
0 & \I_g\\
-\I_g & 0
\end{pmatrix}
\end{align*}
 holds.
	The group is generated by the elements 
\begin{align*}
S = \begin{pmatrix}
A & 0 \\
0 & (A^t  )^{-1}
\end{pmatrix} \textrm{and  } T = \begin{pmatrix}
\I_g & B\\
0 & \I_g
\end{pmatrix}
\end{align*}
that act on the period matrix $\OO \in \mathcal{H}_g$ as follows:  $S: \OO \mapsto A \OO A^t$ and $T: \OO \mapsto \OO + B$. 

The action of the modular group on the pair $(\uz, \OO) \in \C^g \times \mathcal{H}_g$ is \cite{Mumford1983}: $(\uz, \OO) \mapsto (((C\OO +D)^t)^{-1},(A \OO + B)(C\OO + D)^{-1})$, that allows to define the transformation law for $\vartheta$ functions for a generic $\mathcal{R} \in Sp(2g,\R)$:
\begin{align}
\mathcal{R}[\vartheta(\uz|\OO)] = \vartheta\left(((C\OO +D)^t)^{-1} \uz \, | \, (A \OO + B)(C\OO + D)^{-1} \right).
\end{align}

Obviously $\mathcal{R}$ can be specified as the inversion symmetry $\mathcal{P}$, in which case the action reduces to $\mathcal{P}[\vartheta(\uz|\OO)] = \vartheta(-\uz|\OO)$. 
	From the series expansion in Sec. 4.1 in the main text one has that $\vartheta(-\uz|\OO) = \vartheta(\uz|\OO)$, whereas for $\vartheta$ functions with characteristics a simple calculation\cite{Maggio2025} shows that $\ratthe \uw \uk (-\uz|\OO) = \ratthe {-\uw} {-\uk} (\uz|\OO)$.
	
	The action of the inversion symmetry can be evaluated on the Bloch states in the main text, for which an explicit definition involving all coordinates in the Wyckoff position is: $\phi_{\uk}(\uc|\{\uw_\alpha\},\OO) = \sum_{i=1}^4 \phi_{\uk}(\uc|\uw^{(i)}_\alpha,\OO)$, with $i$ running over the elements in the orbit $\{ \uw_\alpha \}$. 
	Starting with the case $\alpha =c$, we have that 
	\begin{align*}
	\mathcal{P} \left[ \phi_{\uk}(\uc|\{\uw_c\},\OO)\right] &= \sum_{i=1}^4 \phi_{\uk}(-\uc|\uw_c^{(i)},\OO)= \sum_{i=1}^4 e^{\ii \pi \uc \cdot \OO \uc} \, e^{-2\ii \pi \uw_c^{(i)} \cdot \uk} \,  \ratthe{\uw_c^{(i)}}{\uk}(-\uz| \OO)\\
	& = \sum_{i=1}^4 e^{\ii \pi \uc \cdot \OO \uc} \, e^{-2\ii \pi \uw_c^{(i)} \cdot \uk} \, \ratthe{-\uw_c^{(i)}}{-\uk}(\uz| \OO)\\
	&= \sum_{i=1}^4 e^{\ii \pi \uc \cdot \OO \uc} \, e^{-2\ii \pi \uw_c^{(i)} \cdot \uk} \, e^{-4\ii \pi \uw_d^{(j)} \cdot \uk} \, \ratthe{\uw_d^{(j)}}{\uk}(\uz| \OO)
	\end{align*}
	where the wavevector coordinates appear as a  characteristic of the $\vartheta$ function, thus $\uz = - \OO \uc$; in going to the last line of the equation above the correspondence between coordinates in $\{ \uw_c\}$ and $\{ \uw_d\}$ under inversion has been exploited with $j=j(i)=2,1,4,3$ for $i=1,2,3,4$ respectively, for the coordinates $\uw_\alpha^{(i)}$ listed in Fig. 1 in the main text.
	The sum above can simply be recast as:
\begin{align}
\label{eq:ParityC}
		\mathcal{P} \left[ \phi_{\uk}(\uc|\{\uw_c\},\OO)\right] &= \sum_{i=1}^4 e^{-2\ii \pi \left( \uw_c^{(i)} + \uw_d^{(j)}\right)\cdot \uk} \phi_{\uk}(\uc|\uw_d^{(j)},\OO) = \phi_{\uk}(\uc|\{ \uw_d\},\OO)
\end{align}	 
since the sum $\uw_c^{(i)} + \uw_d^{(j)} = \un $, with $\un$ integer lattice translation for all $i, j(i)$.
	For the pair $(a,b)$ the derivation is straightforward since the lattices generated by $\{\uw_a\}$ and $\{\uw_b\}$ are invariant under inversion, one thus has:
	\begin{align}
	\label{eq:ParityA}
	\mathcal{P} \left[ \phi_{\uk}(\uc|\{\uw_a\},\OO)\right] = \sum_{i=1}^4 (-1)^{4 \uw_a^{(i)}\cdot \uk} \, \phi_{\uk}(\uc|\uw_a^{(i)},\OO).
	\end{align}
	The parity eigenvalue introduced in the main text, takes values $e_*(\uw_a^{(i)},\Gamma) = e_*(\uw_a^{(i)},Z) =1$ for all $i$ and $e_*(\uw_a^{(i)},\tfrac{1}{2} Z) =\pm 1$ for $i \in \{1,2 \}$ or $i \in \{ 3,4 \}$ respectively (parity need not be conserved along the dispersion curve, since it is not a space group element).
	By introducing the following coefficients: $c_1=[1,1,1,1]; \, c_2=[1,1,-1,-1];\, c_3=[1,-1,1,-1]; \, c_4=[1,-1,-1,1]$, for the linear combinations of BSs $\phi^{(c_n)}_{\uk}(\uc|\{\uw_\alpha\},\OO)=\sum_i c_n(i)\, \phi_{\uk}(\uc|\uw_a^{(i)},\OO)$, one has  that along the dispersion $\Gamma - Z$ (taking the midpoint as representative wavevector) the statements in Eqs. \ref{eq:ParityC} and \ref{eq:ParityA} can be recast as:
	\begin{align*}
	\mathcal{P}\left[ \phi^{(c_n)}_{\uk}(\uc|\{\uw_\alpha\},\OO)\right] & = \phi^{(c_n)}_{\uk}(\uc|\{\uw_\beta\},\OO) \; \; \textrm{for all} \, n \; \; \textrm{and with} \; \; \alpha=c,d \; \; \textrm{and} \; \; \beta=d,c \; \; \textrm{, respectively,} \\
	\mathcal{P}\left[ \phi^{(c_n)}_{\uk}(\uc|\{\uw_\alpha\},\OO)\right] & = \phi^{(c_{j})}_{\uk}(\uc|\{\uw_\alpha\},\OO) \; \; \textrm{for } \,\alpha=a,b  \; \; \textrm{and with  } j(n) = 2, 1, 4, 3 \; \; \textrm{for } n=1, 2, 3, 4 .
	\end{align*}
	So, in practice the inversion symmetry swaps the sublattices $\{\uw_c\} \leftrightarrow \{\uw_d \}$ for all $c_n$, whereas for the BSs @ $\{\uw_a\} , \{ \uw_b\}$ the action keeps the sublattices fixed but swaps the coefficients  $c_n \leftrightarrow c_{n+1}$ for $n=1,3$. 
	This observation can then help rationalise the transformation properties of the lower symmetry BSs under space group operations at the $\Gamma-$point reported in Tab. \ref{tbl:BSIrreps}: the correspondence between WPs $c$ and $d$ implies that BSs have to have matching symmetry properties, which can be different if the overall symmetry is altered by changing the coefficients $c_n$.
	On the other hand, the BSs @ $\{\uw_a\} , \{ \uw_b\}$ do not show symmetry lowering at the hand of the coefficients $c_n$, and the full invariance with respect to space group symmetry operations is retained.

%%%%%%%%%%%%%%%%%%%%%%%%%%%%%%%%%%%%
%%Table 1
%%%%%%%%%%%%%%%%%%%%%%%%%%%%%%%%%%%%
\begin{table}
  \caption{Transformation symmetries for the Bloch states generated by the linear combinations $c_i, \, i=1\dots 4$ of $s$-GTOs located at the Wyckoff positions $\uw_\alpha, \, \alpha=a,b,c,d$.}
  \label{tbl:BSIrreps}
  \begin{tabular}{ccccc}
    \hline
    BS @  & $\Gamma_1$ & $\Gamma_2$ & $\Gamma_3$ & $\Gamma_4$  \\
    \hline
    $\{ \uw_a \}$ & $c_1, \, c_2, \, c_3, \, c_4$   & - & -   & -   \\
    $\{ \uw_b \}$ & $c_1, \, c_2, \, c_3, \, c_4$ & - & -   & - \\
    $\{ \uw_c \}$ & $c_1$  & $c_3$ & $c_2$   & $c_4$ \\
    $\{ \uw_d \}$ & $c_1$  & $c_3$ & $c_2$   & $c_4$ \\
    \hline
  \end{tabular}
\end{table}
%%%%%%%%%%%%%%%%%%%%%%%%%%%%%%%%%%%%%%%%%%

\section{Character table and additional plots for the BSs}

\begin{figure}[h]
\includegraphics[width=180mm]{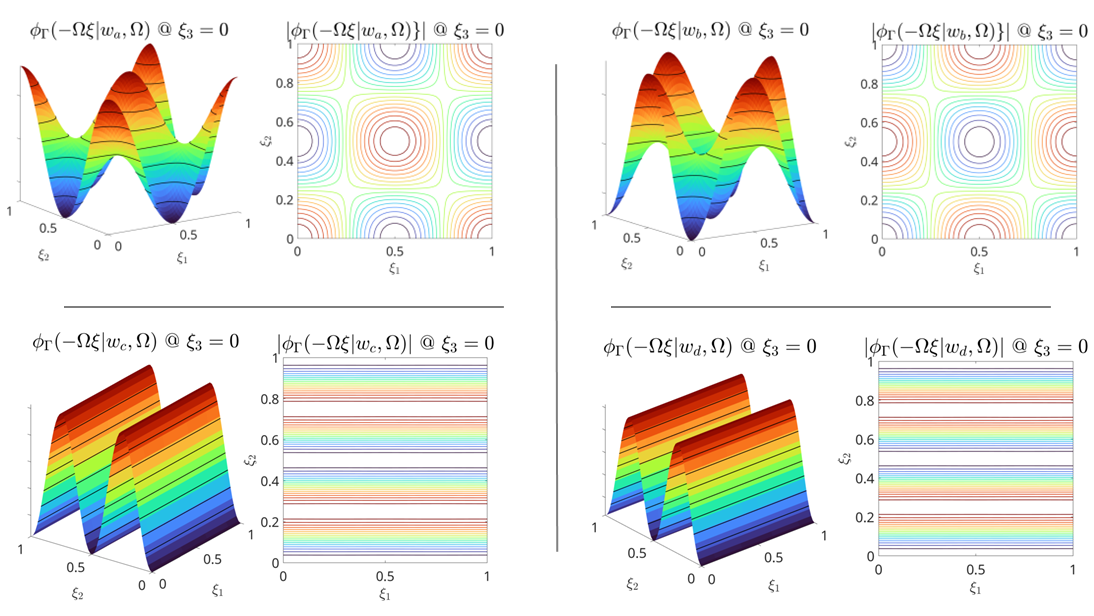}
\caption{Bloch states @ $\{ \uw_\alpha\}, \, \alpha=a,b,c,d$. Each panel shows the real part (left) and the absolute value of the BS in the plane $\xi_3 = 0$.}
\label{fig:BSsXi3}
\end{figure}

Character table for $G_\Gamma$, the little group at the $\Gamma$-point (Tbl.\ref{tbl:CC_Gamma}) and section in the plane $\xi_3=0$ of the BSs considered in the main text (Fig. \ref{fig:BSsXi3}).

\begin{table}
\caption{Character table for $G_\Gamma$, the little group at the $\Gamma$-point}
\label{tbl:CC_Gamma}
\begin{tabular}{c|cccc}
\hline
Irrep & $e$ & $m_x$ & $m_z$ & $m_y$ \\
\hline
$\Gamma_1$ & $1$ & $\phantom{-}1 $ & $ \phantom{-}1 $ & $ \phantom{-}1 $ \\
$\Gamma_2$ & $1$ & $-1 $ & $-1 $ & $ \phantom{-}1 $ \\
$\Gamma_3$ & $1$ & $-1 $ & $ \phantom{-}1 $ & $-1 $ \\
$\Gamma_4$ & $1$ & $ \phantom{-}1 $ & $-1 $ & $-1 $ \\
\end{tabular}
\end{table}

\section{Additional derivations}
\subsection{Derivative of Bloch state's periodic component}
\label{sec:BSderv}
The derivative of the unnormalised periodic component of the Bloch state, $\tilde{u}_{\uk}$, is reported in Sec. \ref{ssec:dudk} and it evaluates as: 
\begin{align*}
\frac{\partial}{\partial k^\mu} \tilde{u}_{\uk}(\uc|\uw,\OO) = \frac{\partial}{\partial k^\mu} \left( e^{-2\ii \pi \uc \cdot \uk} \, \tilde{\phi}_{\uk}(\uc|\uw,\OO)\right)= -2\ii \pi (\ue^\mu,\uc+\uw)e^{-2\ii \pi \uc \cdot \uk} \, \tilde{\phi}_{\uk}(\uc|\uw,\OO) + e^{-2\ii \pi \uc \cdot \uk} \, \tilde{\phi}'_{\ue^\mu,\uk}(\uc|\uw,\OO)
\end{align*}
which can then be replaced in the corresponding expression for the normalised $u_{\uk}$:
\begin{align*}
\frac{\partial}{\partial k^\mu} u_{\uk}(\uc|\uw,\OO)= \left( \frac{\partial}{\partial k^\mu} N_{\uk}\right)\tilde{u}_{\uk}(\uc|\uw,\OO) + N_{\uk} \frac{\partial}{\partial k^\mu} \tilde{u}_{\uk}(\uc|\uw,\OO)
\end{align*}
with the assumption leading to Eq. 3 in the main text, the derivative above reads:
\begin{align*}
\frac{\partial}{\partial k^\mu} u_{\uk}(\uc|\uw,\tau \dd)=& \left \{ - \left[ 2\ii \pi (\ue^\mu, \uc + \uw) + \frac{1}{2} \frac{\vartheta'(k^\mu|\oo^\mu)}{\vartheta(k^\mu|\oo^\mu)} \right] \phi_{\uk}(\uc|\uw,\tau \dd) +\right.\\
\left. \phi'_{\ue^\mu,\uk}(\uc|\uw,\tau \dd) \right\} 
\end{align*}
where $\oo^\mu=\tfrac{1}{2}\tau (\ue^\mu, \dd)$ is the component of the period matrix along the direction in which the derivative is evaluated.

	Next I am reporting the evaluation of the matrix element of the Berry connection, given in Eq. 4 in the main text.
	\begin{align}
	\left\langle u_{\uk} \bigg| \frac{\partial}{\partial k^\mu} u_{\uk}\right\rangle =& N_{\uk}^2 \int_{\U} d\uc \, \tilde{\phi}_{-\uk}(\uc|\uw,\tau \dd)\, (-2\pi \ii) \, (\ue^\mu,\uc+\uw) \, \tilde{\phi}_{\uk}(\uc|\uw,\tau \dd) \label{eq:1stLine}\\
	& - \tfrac{1}{2} N_{\uk}^2 \frac{\vartheta'(k^\mu|\oo)}{\vartheta(k^\mu|\oo)} \, \int_{\U}d\uc \, \tilde{\phi}_{-\uk}(\uc|\uw,\tau \dd)\, \tilde{\phi}_{\uk}(\uc|\uw,\tau \dd)\label{eq:2ndLine}\\
	& + N_{\uk}^2 \int_{\U} d\uc \, \tilde{\phi}_{-\uk}(\uc|\uw,\tau \dd)\, \tilde{\phi}'_{\ue^\mu,\uk}(\uc|\uw,\tau \dd). \label{eq:3rdLine}
	\end{align}
	The integral over the unit cell $\U$ in Eq.\ref{eq:2ndLine} above evaluates to:
	\begin{align}
	\int_{\U}d\uc \, \tilde{\phi}_{-\uk}(\uc|\uw,\tau \dd)\, \tilde{\phi}_{\uk}(\uc|\uw,\tau \dd) = (abc)^{-1}\left(\frac{\pi}{2\beta}\right)^{3/2} \prod_{\mu=1}^3\vartheta(k^\mu|\oo^\mu),
	\end{align}
 whereas the sum of the remaining terms in Eqs. \ref{eq:1stLine}, \ref{eq:3rdLine} is:
 \begin{align}
 \label{eq:sum4-6}
 N_{\uk}^2 \int_{\U} d\uc \, \tilde{\phi}_{-\uk}(\uc|\uw,\tau \dd)\, \left[ \tilde{\phi}'_{\ue^\mu,\uk}(\uc|\uw,\tau \dd) - 2\pi \ii (\ue^\mu,\uc+\uw) \tilde{\phi}_{\uk}(\uc|\uw,\tau \dd)\right]
 \end{align}
 which can be evaluated straightforwardly if one recognises the expression for the $p$-GTO BS \cite{Maggio2025}: $-2\pi\ii \, \tilde{\phi}_{\uk}^{(p_\mu)}=\tilde{\phi}'_{\ue^\mu,\uk} -2\pi\ii (\ue^\mu,\uc) \, \tilde{\phi}_{\uk} $.
 Then one has that the term in Eq. \ref{eq:sum4-6} above evaluates as:
 \begin{align*}
 N_{\uk}^2 \int_{\U}d\uc \, \tilde{\phi}_{-\uk}(\uc|\uw,\tau \dd)\, \tilde{\phi}_{\uk}^{(p_\mu)}(\uc|\uw,\tau \dd) \\
 - 2\pi \ii N_{\uk}^2 \, (\ue^\mu,\uw)\int_{\U}d\uc \, \tilde{\phi}_{-\uk}(\uc|\uw,\tau\dd)\, \tilde{\phi}_{\uk}(\uc|\uw,\tau\dd)
 \end{align*}
 
 Recalling that (see Sec. \ref{sec:orth} below) $\langle \tilde{\phi}_{\uk}|\tilde{\phi}_{\uk}^{(p_\mu)}\rangle=0$ one obtains:
 \begin{align*}
 \left\langle u_{\uk} \bigg| \frac{\partial}{\partial k^\mu} u_{\uk}\right\rangle = -\tfrac{1}{2} (abc)^{-1} \left(\frac{\pi}{2\beta}\right)^{3/2} \, N_{\uk}^2 \, \vartheta(k^\nu|\oo^\nu) \, \vartheta(k^\lambda|\oo^\lambda) \left[ \vartheta'(k^\mu|\oo^\mu) + 4\pi\ii (\ue^\mu,\uw)\, \vartheta(k^\mu|\oo^\mu)\right]
 \end{align*}
 By replacing the expression for the normalisation constant: $N_{\uk}^2 = (abc)\left(\frac{2\beta}{\pi}\right)^{3/2} [\, \vartheta(k^\mu|\oo^\mu)\, \vartheta(k^\nu|\oo^\nu)\, \vartheta(k^\lambda|\oo^\lambda)]^{-1}$, Eq. 4 in the main text is obtained.

\subsection{Derivative of the periodic component of the Bloch states}
\label{ssec:dudk}
	In this section I report the derivation of the equality:
\begin{align}
\frac{\partial}{\partial k^\mu} \tilde{u}_{\uk}(\uc|\uw,\OO) = \frac{\partial}{\partial k^\mu} \left( e^{-2\ii \pi \uc \cdot \uk} \, \tilde{\phi}_{\uk}(\uc|\uw,\OO)\right)= -2\ii \pi (\ue^\mu,\uc+\uw)e^{-2\ii \pi \uc \cdot \uk} \, \tilde{\phi}_{\uk}(\uc|\uw,\OO) + e^{-2\ii \pi \uc \cdot \uk} \, \tilde{\phi}'_{\ue^\mu,\uk}(\uc|\uw,\OO).
\end{align}
The statement can be derived easily by applying the following:
\begin{align}
\label{eq:dphi}
\frac{\partial}{\partial k^\mu} \tilde{\phi}_{\uk}(\uc|\uw,\OO) = \tilde{\phi}'_{\ue^\mu,\uk}(\uc-\uw|\uzz, \OO),
\end{align}
and the translational invariance reported in Eq.(10) of ref. \cite{Maggio2025}:
\begin{align*}
\tilde{\phi}'_{\ue,\uk}(\uc|\uw,\OO)=2\pi\ii(\ue,\uw)\tilde{\phi}_{\uk}(\uc|\uw,\OO)+\tilde{\phi}'_{\ue,\uk}(\uc-\uw|\uzz,\OO).
\end{align*}

	Eq.\ref{eq:dphi} can then be immediately obtained as follows:
	\begin{align*}
	\frac{\partial}{\partial k^\mu} \tilde{\phi}_{\uk}(\uc|\uw,\OO) & = e^{\ii \pi \uc \cdot \OO \uc} \, e^{-2\pi\ii \uw \cdot \uk} \left( D_{\ue^\mu} \left[ \ratthe{\uw}{\uzz}(\uk -\OO \uc|\OO)\right] -2 \pi \ii (\ue^\mu,\uw) \, \ratthe{\uw}{\uzz}(\uk-\OO \uc|\OO)  \right)\\
	& = \tilde{\phi}'_{\ue^\mu,\uk}(\uc|\uw,\OO) - 2\pi\ii (\ue^\mu, \uw) \tilde{\phi}_{\uk}(\uc|\uw,\OO),
	\end{align*}
and by applying the chain rule for the derivatives.

\subsection{Bloch states from s- and p- GTOs are orthogonal}
\label{sec:orth}
The expression for the overlap integral of $s$- and $p$-BSs in one cartesian dimension reads, by setting the lattice constant to 1, as:
\begin{align*}
S_k^{(0,1)}=\langle \tilde{\phi}_k^{(0)} | \tilde{\phi}_k^{(1)} \rangle = \int_0^1 dx \tilde{\phi}_{-k}^{(0)}(x) \left[ x \, \tilde{\phi}_k^{(0)}(x) - \tfrac{1}{2\ii \pi} \tfrac{d}{dx} \tilde{\phi}_k^{(0)}(x)\right],
\end{align*}
where the  $s$- and $p$- Bloch states have been indicated by the respective quantum number and the expression for the $\tilde{\phi}_k^{(1)}$ has been given in Eq. 12 of ref. \cite{Maggio2025} . 
	The dependence on the Wyckoff position has been omitted, since the integral is insensitive to it, as long as states arise from GTOs on the same Wyckoff position.
	By plugging in the appropriate definitions, the integral becomes:
	\begin{align}
	S_k^{(0,1)} &= \sum_{m,n \in \Z} e^{2\ii \pi k(n-m)} \, e^{-\tfrac{\beta}{2}(n-m)^2} \, \int_0^1 dx \, (x-n) e^{-2\beta \left(x - \tfrac{1}{2}(m+n) \right)^2} \\
	&= \sum_{s,n \in \Z} e^{-2\ii \pi ks} \, e^{-\tfrac{\beta}{2}(-s)^2} \, \int_0^1 dx \, (x-n) e^{-2\beta \left(x -\tfrac{s}{2} -n \right)^2}.
	\end{align}
	By means of a change of variable, it is possible to carry out the summation over the $n$ index to obtain:
	\begin{align}
	S_k^{(0,1)} = \sum_{s \in \Z} e^{-2\ii \pi ks} \, e^{-\tfrac{\beta}{2}(-s)^2} \, \int_{-\infty}^{+\infty} dy \, y \, e^{-2\beta \left(y-\tfrac{s}{2} \right)^2} =0,
	\end{align}
	where the last equality follows on the ground of parity.

%\begin{figure}
%  As well as the standard float types \texttt{table}\\
%  and \texttt{figure}, the class also recognises\\
%  \texttt{scheme}, \texttt{chart} and \texttt{graph}.
%  \caption{An example figure}
%  \label{fgr:example}
%\end{figure}

%\begin{table}
%  \caption{An example table}
%  \label{tbl:example}
%  \begin{tabular}{ll}
%    \hline
%    Header one  & Header two  \\
%    \hline
%    Entry one   & Entry two   \\
%    Entry three & Entry four  \\
%    Entry five  & Entry five  \\
%    Entry seven & Entry eight \\
%    \hline
%  \end{tabular}
%\end{table}

\bibliography{MyArticles, Books}